\documentclass[conference]{IEEEtran}
\pdfoutput=1
\usepackage{amsmath}
\usepackage{algorithmic}
\usepackage{graphicx}
\usepackage{svg}
\usepackage[skip=3pt]{caption}
\usepackage{array}
\usepackage{cite}
\usepackage{hyperref}
\usepackage{xcolor,soul}
\usepackage{amssymb}

\usepackage{booktabs, makecell, multirow, tabularx} 
\newcolumntype{C}{>{\centering\arraybackslash}X}
\newcolumntype{L}{>{\hsize=.4\hsize}C}
\newcolumntype{M}{>{\hsize=.35\hsize}C}
\newcolumntype{S}{>{\hsize=.25\hsize}C}
\newcommand{\mcccll}[1]{\multicolumn{3}{|c||}{#1}}
\newcommand{\mcccl}[1]{\multicolumn{3}{c|}{#1}}
\newcommand{\mccl}[1]{\multicolumn{2}{|c|}{#1}}

\usepackage{colortbl}
\definecolor{light-gray}{gray}{0.85}
\usepackage{caption} 
\usepackage{diagbox}
\usepackage{threeparttable}

\usepackage[ruled, linesnumbered, vlined]{algorithm2e}
\SetAlFnt{\small}
\SetAlgoNlRelativeSize{0}
\IncMargin{3pt}
\SetKwInOut{Input}{Input}
\SetKwInOut{Require}{Require}
\SetKwInOut{Output}{Output}

\newenvironment{para_noindent}{\setlength{\parindent}{0pt}}{}
\definecolor{moh_colour}{RGB}{255, 204, 204}

\definecolor{yuz_colour}{RGB}{191, 232, 255}

\begin{document}

\title{BRAMAC: Compute-in-BRAM Architectures for Multiply-Accumulate on FPGAs}


\author{Yuzong Chen and Mohamed S. Abdelfattah \\
        \textit{Department of Electrical and Computer Engineering, Cornell University}  \\
        \{yc2367, mohamed\}@cornell.edu
        }


\maketitle

\begin{abstract}
Deep neural network (DNN) inference using reduced integer precision has been shown to achieve significant improvements in memory utilization and compute throughput with little or no accuracy loss compared to full-precision floating-point. 
Modern FPGA-based DNN inference relies heavily on the on-chip block RAM (BRAM) for model storage and the digital signal processing (DSP) unit for implementing the multiply-accumulate (MAC) operation, a fundamental DNN primitive. 
In this paper, we enhance the existing BRAM to also compute MAC by proposing BRAMAC (Compute-in-\underline{BR}AM \underline{A}rchitectures for \underline{M}ultiply-\underline{Ac}cumulate). BRAMAC supports 2's complement 2- to 8-bit MAC in a small dummy BRAM array using a hybrid bit-serial \& bit-parallel data flow. 
Unlike previous compute-in-BRAM architectures, BRAMAC allows read/write access to the main BRAM array while computing in the dummy BRAM array, enabling both persistent and tiling-based DNN inference. 
We explore two BRAMAC variants: BRAMAC-2SA (with 2 synchronous dummy arrays) and BRAMAC-1DA (with 1 double-pumped dummy array). 
BRAMAC-2SA/BRAMAC-1DA can boost the peak MAC throughput of a large Arria-10 FPGA by 2.6$\times$/2.1$\times$, 2.3$\times$/2.0$\times$, and 1.9$\times$/1.7$\times$ for 2-bit, 4-bit, and 8-bit precisions, respectively at the cost of 6.8\%/3.4\% increase in the FPGA core area. 
By adding BRAMAC-2SA/BRAMAC-1DA to a state-of-the-art tiling-based DNN accelerator, an average speedup of 2.05$\times$/1.7$\times$ and 1.33$\times$/1.52$\times$ can be achieved for AlexNet and ResNet-34, respectively across different model precisions. Our code is available at: \url{https://github.com/abdelfattah-lab/BRAMAC}.
\end{abstract}


\section{Introduction}
Deep neural networks (DNNs) have become ubiquitous in many important fields such as computer vision, speech recognition, and natural language processing. 
However, a well-trained DNN model for complicated tasks has a huge model size ranging from several hundreds of megabytes (e.g., AlexNet classifying ImageNet) to several hundreds of gigabytes (e.g. GPT3 producing human-like text) \cite{AlexNet, ImageNet, GPT3}. 
Accordingly, many researchers have been exploring reduced numerical precisions to represent DNN model weights and activations, especially during inference where reduced-precision arithmetic incurs little or no accuracy loss compared to full-precision floating-point (FP) \cite{NNQuantization, IntegerQuantization}. 
This low-precision property allows better utilization of on-chip memory and computation resources for improved performance. 
For example, Nvidia GPUs can obtain a 4-8$\times$ inference speedup using INT8 precision compared to FP32 precision \cite{NvidiaInt8}, and an additional 1.6$\times$ speedup using INT4 precision compared to INT8 \cite{NvidiaInt4}. 
    
In the meanwhile, FPGAs are becoming an increasingly popular platform for DNN acceleration due to their hardware programmability that enables customized datapaths and numerical bit-widths suitable for low-precision inference \cite{BrainWave, DLA_old, DLA_new, IntelStratix10NX}. 
FPGA-based DNN accelerators heavily rely on block random access memory (BRAM) for model storage and digital signal processing (DSP) units for implementing multiply-accumulate (MAC)---the fundamental primitive in DNNs. 
Nevertheless, most FPGA vendors' DSP blocks do not natively support precisions lower than 18 bits, making them sub-optimal for implementing low-precision MAC \cite{XilinxDSP, IntelDSP, AchronixlDSP}. 
For DNNs to better utilize FPGA's on-chip resources, researchers have proposed novel DSP architectures for low-precision MAC \cite{EmbracingDiversity, PIR_DSP}. 
More recently, some works have proposed to add compute capability inside BRAMs and enable them to perform various Boolean and arithmetic operations \cite{CCB, CoMeFa}. 
This computing in-memory (CIM) approach does not sacrifice the performance of existing logic resources on FPGA but rather complements them to further boost the FPGA's computing throughput. 
In addition, CIM can reduce the routing associated with data movement between memory and logic units, hence saving energy and area. 
This is especially true in DNN accelerators where model parameters and activations are frequently transferred between BRAMs and DSPs to perform massive computations.

In this paper, we further enhance the FPGA's compatibility with low-precision DNNs by proposing BRAMAC, an efficient compute-in-BRAM architecture for multiply-accumulate. 
Unlike previous CIM architectures that compute directly on the main BRAM array \cite{CoMeFa, CCB}, 
BRAMAC first copies the data from the main BRAM array to an additional, separate memory array and then computes on this ``dummy" array, which is a true dual-port BRAM with the same number of columns as the main BRAM array but only 7 rows. 
This 7-row dummy array can be accessed fast with low power consumption due to a much smaller parasitic load on its bitlines compared to the main BRAM array which typically has $>$100 physical rows. 
Furthermore, the dummy array allows BRAMAC to function like a normal BRAM even during CIM operations---the main BRAM array's read and write ports are available for use by the application logic.
Finally, BRAMAC is optimized for DNN MAC operations by performing shared-input multiplication and in-place accumulation.
We enumerate our contributions below:
\begin{enumerate}
    \item We propose new peripheral circuits that enable BRAMAC to compute two MACs (or one MAC2), $P = (W_{1}I_{1} + W_{2}I_{2})$, simultaneously using a hybrid bit-serial \& bit-parallel dataflow.
    \item We propose two BRAMAC variants with different area-throughput trade-offs: BRAMAC with 2 synchronous dummy arrays (2SA) and BRAMAC with one double-pumped dummy array (1DA).
    \item We design an embedded finite-state machine (eFSM) to free up the main BRAM ports during MAC2 computation and to allow simultaneous main BRAM access, thus enabling efficient tiling-based DNN acceleration. 
    \item We quantify the benefits of employing BRAMAC in a tiled FPGA DNN accelerator, which achieves up to 2.04$\times$ and 1.52$\times$ performance improvements for AlexNet and ResNet-34, respectively over the baseline accelerator without BRAMAC.
\end{enumerate}


\section{Related Work} \label{Related_Works}
In this section. we discuss previous work that targeted efficient MAC implementation on FPGAs including logic block, DSP, and BRAM enhancements.
    
\subsection{Logic Block with Fast Arithmetic}
To efficiently implement arithmetic operations in soft logic, modern FPGAs contain hardened adder circuitry in their logic blocks (LBs) \cite{FPGA_Survery}. These adders range from simple ripple-carry adders to more complex variants such as carry-bypass adders and carry-lookahead adders. In order to reduce the carry propagation delay, dedicated routing is used to propagate carry signals between different LBs. 
Inspired by the superior efficiency of adopting low-precision in DNN, recent research started to investigate adding more hardened arithmetic in LBs. 
For example, Boutros~\textit{et al.} \cite{LB_MAC} proposed three LB architectural enhancements to improve the performance of MAC implemented in soft logic. Their most promising proposal increases the MAC density by 1.7$\times$ while simultaneously improving the MAC speed. 

\subsection{Low-Precision DSP Architectures}
Modern commercial FPGAs include DSP blocks that implement efficient multiplication with additional features such as pre-addition and accumulation commonly used in signal processing applications \cite{FPGA_Survery}. Nevertheless, most FPGA vendors' DSP multipliers have a minimum precision of 18-bit, making them less competitive in accelerating low-precision DNNs. To address this limitation, researchers have proposed new DSP architectures to support low-precision MAC. Boutros \textit{et al.} \cite{EmbracingDiversity} introduced an enhanced Intel DSP (eDSP) that supports four 9-bit or eight 4-bit multiplications without using additional routing ports. Rasoulinezhad \textit{et al.} \cite{PIR_DSP} presented a modified Xilinx DSP, called PIR-DSP, that can carry out six 9-bit, twelve 4-bit, or twenty-four 2-bit multiplications. Regarding industry DSP trends, the recent Xilinx Versal and Intel Agilex devices added support for 8-bit multiplication in their DSP blocks \cite{VersalDSP, AgilexDSP}. In addition, Intel's latest Stratix-10 NX device added a new DSP (called AI tensor) block that contains 30 INT8 multipliers and can also be configured as 60 INT4 multipliers \cite{stratix10nx}.
    

\subsection{Computing In-BRAM}
With the emergence of CIM to overcome the von-Neumann bottleneck \cite{ComputingEnergy}, some FPGA researchers suggest augmenting existing BRAM architectures with compute capability. 
Wang \textit{et al.} \cite{CCB} proposed a compute-capable BRAM (CCB) that uses bit-serial arithmetic to enable a high degree of computation parallelism. 
However, the circuit implementation of CCB requires an additional voltage supply to mitigate the read-disturb issue associated with activating two word-lines from one BRAM port, which is challenging to implement in practice. 
Arora \textit{et al.} \cite{CoMeFa} later designed a new compute-in-BRAM architecture called CoMeFa to overcome some limitations of CCB. 
CoMeFa also relies on bit-serial arithmetic but exploits the dual-port nature of BRAM to read out two operands from two ports, respectively instead of activating two word-lines from one port, thus eliminating the read-disturb issue.

Both CCB and CoMeFa require transposed data layout for bit-serial computation, i.e., each word occupies one column and multiple rows instead of one row and multiple columns in a conventional data layout. However, transposing data is expensive in both latency and additional hardware cost (e.g. a swizzle module in CoMeFa) for online execution. 
Furthermore, these two BRAM architectures compute directly on the main BRAM array and receive the CIM instruction through a BRAM write port---this prevents tiling.
As a result, these two works are limited to accelerating only persistent-style DNN inference where the model weights are transposed offline and remain persistent in the on-chip memory.
Different from CCB and CoMeFa, BRAMAC adopts a hybrid bit-serial \& bit-parallel MAC dataflow that eliminates the requirement of transposed data layout. In addition, BRAMAC doesn't compute on the main BRAM array which is typically large and therefore, slow and power-hungry. Rather, it copies the main BRAM's data to a special, separate dummy BRAM array for computation. 
This dummy array has only 7 rows and therefore can be accessed much faster compared to the main BRAM array.
It can also free up the read and write ports of the main BRAM during CIM to allow tiling-based DNN acceleration.

\section{BRAMAC Architecture and Dataflow} \label{BRAMAC_Architecture}

\subsection{Overall Architecture}
\begin{figure}
    \centering
    \includegraphics[width=1\linewidth]{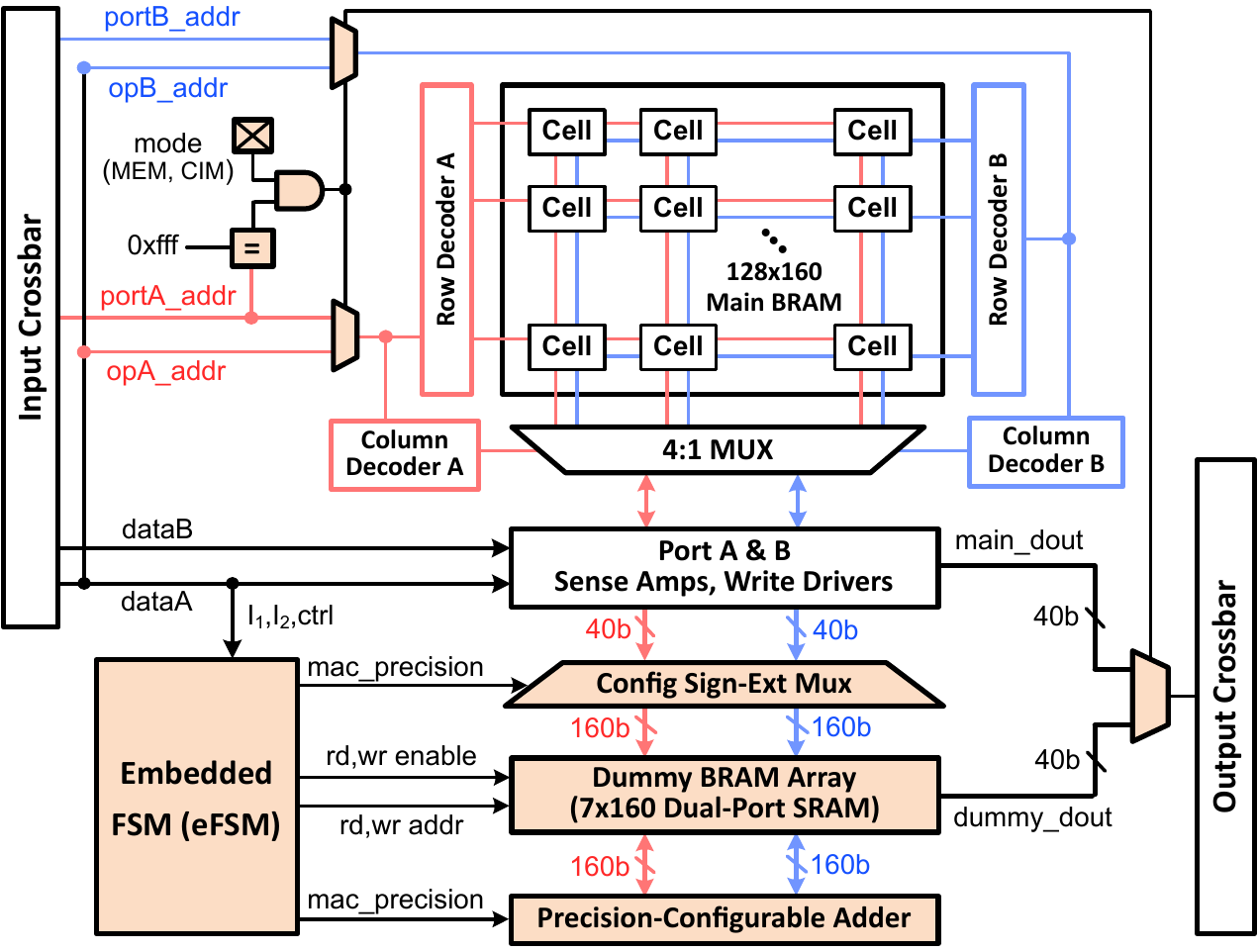}
    \caption{Top-level block diagram of BRAMAC modified from Intel's M20K BRAM. New circuit blocks are orange-shaded.}
    \label{fig1_bramac_architecture}
\end{figure}

Fig. \ref{fig1_bramac_architecture} shows the top-level block diagram of BRAMAC modified from Intel's M20K BRAM \cite{M20K} with added circuit blocks orange-shaded. The routing interface (i.e., input and output crossbar) of BRAMAC is the same as that of M20K. The main BRAM array's dimension is 128-row $\times$ 160-column, i.e., 20 kb memory capacity. The 4:1 column multiplexing feature of M20K is preserved. One additional SRAM cell is added to select one of the two operation modes of BRAMAC:

\subsubsection{MEM}
In this memory mode, the behavior of BRAMAC is identical to that of a conventional M20K.
The input crossbar sends the address and data to portA and portB. 
For memory reads, the two addresses are decoded by the row and column decoders. The 40-bit BRAM output data from sense amplifiers is sent to the output crossbar. 
For memory writes, the data is sent to the write drivers for updating the main BRAM. 
    
\subsubsection{CIM} \label{bramac_mode_cim}
This is the compute mode where BRAMAC can compute MAC2, $P = (W_{1}I_{1} + W_{2}I_{2})$, using 2-bit, 4-bit, or 8-bit operand precision. The two groups of operands, $(W_{1}, W_{2})$ and $(I_{1}, I_{2})$, can be thought of as weights and inputs of DNN in the remainder of this paper, respectively. At a high level, BRAMAC computes MAC2 by keeping weights inside BRAMAC while streaming inputs from outside. 

The main BRAM is automatically configured as a simple dual-port memory with a maximum data width of 40-bit, and a depth of 512 to maximize the read/write throughput. A special address (0xfff) is reserved and compared with the portA address, and if equal, the 40-bit portA data is treated as a CIM instruction.
The CIM instruction contains two addresses for reading two 40-bit data from the main BRAM, respectively. Each 40-bit data is a vector that contains multiple low-precision $W_{1}$/$W_{2}$ elements. The configurable sign-extension mux sign-extends the 40-bit vectors to 160-bit before copying them to a dummy BRAM array which is a 7-row $\times$ 160-column true dual-port BRAM without the column multiplexing feature. The CIM instruction also contains the two inputs, $I_{1}$ and $I_{2}$, and several control signals that are sent to an eFSM to trigger and control the MAC2 operation. The precision-configurable adder can read two 160-bit vectors from the dummy array, performs a single-instruction-multiple-data (SIMD) add, and writes the sum back to the dummy array. 
Since the dummy array has the same number of columns as the main BRAM array, it can read out 40-bit data similar to the main BRAM. A 2-to-1 mux is added to select the data between the main BRAM and the dummy array. 

\subsection{Hybrid Bit-Serial \& Bit-Parallel MAC Dataflow} \label{mac2_dataflow}
BRAMAC computes 2's complement MAC2 by adopting a hybrid bit-serial \& bit-parallel dataflow \cite{stripes} as described in Algorithm \ref{mac2_algorithm}. 
The for-loop in line \ref{mac2algo_beginfor}-\ref{mac2algo_endfor} iterates through two inputs bit-by-bit. 
Each iteration involves multiplying the entire $W_{1}$ and $W_{2}$ by a single bit from $I_{1}$ and $I_{2}$, respectively, followed by a bit-parallel addition to obtain the partial sum (psum) as shown in line \ref{mac2algo_psum}. 
If the current input bit is the most-significant bit (MSB), then psum is subtracted from P (line \ref{mac2algo_minus_psum}) since the MSB is negative in 2's complement representation. 
If the current input bit is not the least-significant bit (LSB), then P also needs to be shifted left by 1-bit after adding psum (lines \ref{mac2algo_shift1_1}, \ref{mac2algo_shift1_2}). 

\begin{algorithm} [t]
    \caption{Hybrid Bit-Serial \& Bit-Parallel MAC2}
    \label{mac2_algorithm}

    \Require{All numbers are integers in 2's complement}
    \Input{$W \in \mathbb{Z}^{2}$, $I \in \mathbb{Z}^{2}$, precision $n \geq 2$}
    \Output{$P \in \mathbb{Z}$}
    
    \textbf{Initialization} $P = 0$ \\ \label{mac2algo_init}
    \For{$i = (n-1)$ downto $0$}{  \label{mac2algo_beginfor}
        $psum = W_{1}*I_{1}[i] + W_{2}*I_{2}[i]$ \\ \label{mac2algo_psum}
        \uIf{$i == (n-1)$} {
            $P = P + inv(psum) +1$ \\ \label{mac2algo_minus_psum}
            $P = P << 1$ \label{mac2algo_shift1_1}
        } \uElseIf {$i \neq 0$} {
            $P = P + psum$ \\ \label{mac2algo_add_psum1}
            $P = P << 1$ \label{mac2algo_shift1_2}
        } \Else {
            $P = P + psum$ \label{mac2algo_add_psum2}
        }
    } \label{mac2algo_endfor}
    \Return{$P$} \\ 
\end{algorithm}

\begin{figure}
    \centering
    \includegraphics[width=1\linewidth]{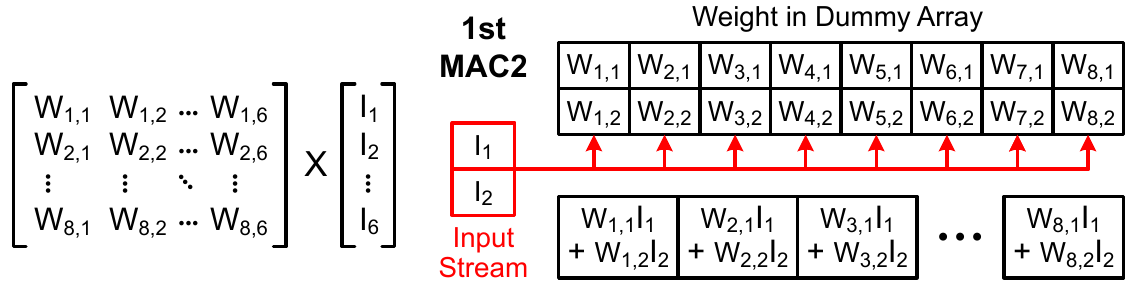}
    \caption{Example of MAC2 to compute matrix-vector multiplication.} 
    \label{fig2_mvm}
\end{figure}

The hybrid bit-serial \& bit-parallel MAC2 algorithm is efficient for computing matrix-vector multiplication (MVM) where the $n^{th}$ vector element is multiplied by all elements of the $n^{th}$ matrix column. 
To exploit this input-sharing in BRAMAC, two inputs are packed into the CIM instruction that is sent to BRAMAC, then multiplied by all elements of the corresponding two matrix columns copied to the dummy array, respectively. 
Copying a matrix column requires the weight matrix to be transposed so that matrix columns correspond to a BRAM row. This can be easily done offline for DNNs. 
Fig. \ref{fig2_mvm} illustrates an example of using MAC2 to compute MVM where the matrix dimension is 8$\times$6. 
For the first MAC2, the first and second matrix columns are copied from the main BRAM to the dummy array. 
Two vector elements $I_{1}$ and $I_{2}$ are streamed to BRAMAC through the CIM instruction and multiplied by all 8 elements of the first and second matrix columns to obtain 8 partial sums. 
For large matrices, the number of matrix elements that can be loaded to the dummy array depends on the MAC precision. 
Since the two read ports of the main BRAM have a total data width of 80-bit, they can copy ten 8-bit, twenty 4-bit, or forty 2-bit weights to a dummy array for one MAC2, providing a parallelism of 10, 20, or 40 MACs, respectively.

\subsection{Circuit Design to Support MAC2}
We now describe the new circuit blocks in BRAMAC to support MAC2. 
These circuit blocks are shown in Fig. \ref{fig3_dummy_array_circuit}, including a dual-port ``dummy" BRAM array (Fig. \ref{fig3_dummy_array_circuit}(a)), a configurable sign-extension mux (Fig. \ref{fig3_dummy_array_circuit}(b)), a 160-bit SIMD adder implemented using 1-bit full adders, and read/write circuits (Fig. \ref{fig3_dummy_array_circuit}(c)).

\subsubsection{Dual-Port Dummy BRAM Array}
\begin{figure} [t]
    \centering
    \includegraphics[width=1\linewidth]{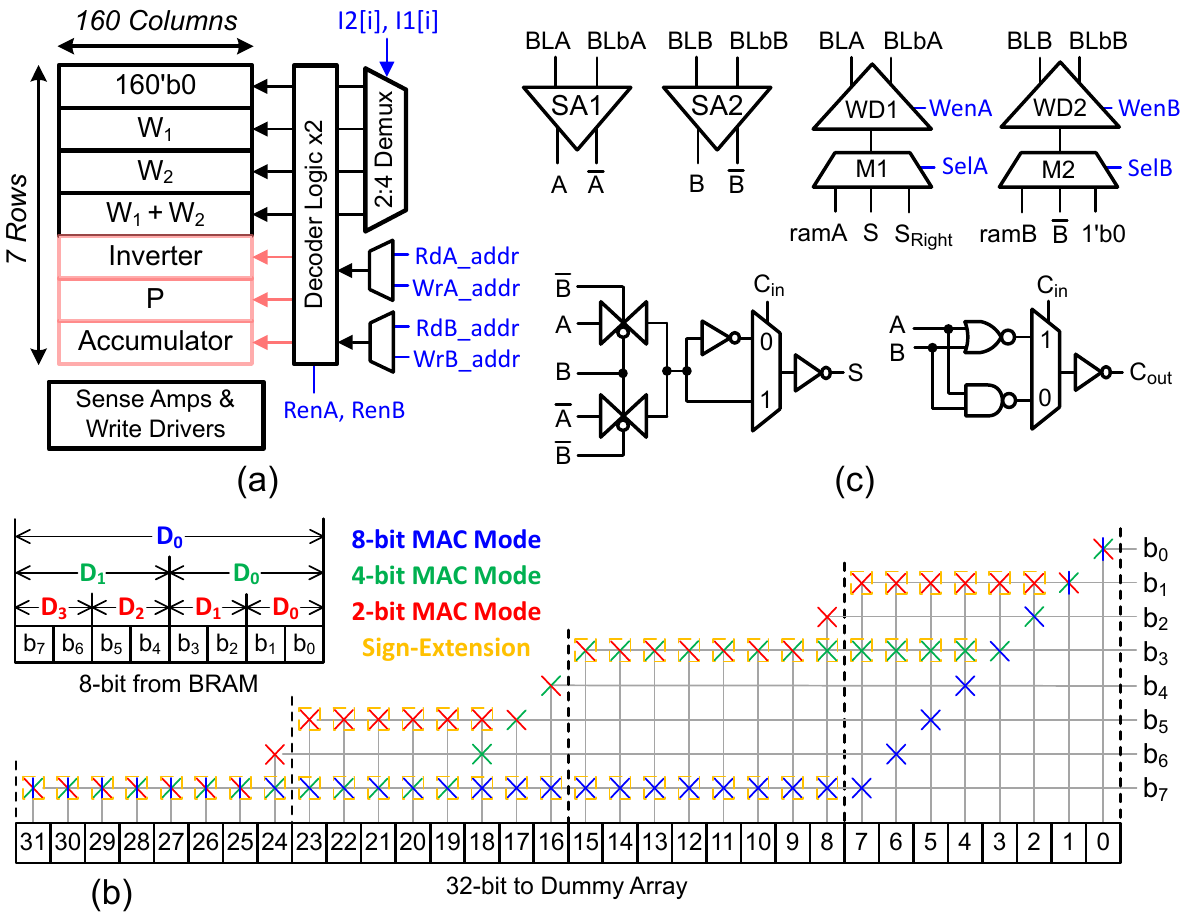}
    \caption{BRAMAC circuit blocks for computing MAC2: (a) dual-port dummy BRAM array, (b) configurable sign-extension mux (here we are displaying one out of five identical blocks), (c) 1-bit full-adder with read/write circuits.}
    \label{fig3_dummy_array_circuit}
\end{figure}
    
The dual-port dummy BRAM array is 7-row $\times$ 160-column without column multiplexing as shown in Fig. \ref{fig3_dummy_array_circuit}(a). 
Its SRAM cell is identical to that used in the main BRAM. 
Each column contains two sense amplifiers and two write drivers to allow true dual-port access. 
Its 1\textsuperscript{st} row is hard-coded to always store \textbf{0}. The 2\textsuperscript{nd} and 3\textsuperscript{rd} rows store the \textbf{W\textsubscript{1}} and \textbf{W\textsubscript{2}} vectors, respectively that are copied from the main BRAM array. 
The 4\textsuperscript{th} row stores a \textbf{(W\textsubscript{1} + W\textsubscript{2})} vector.
The 5\textsuperscript{th} \textbf{Inverter} row is used to store the temporary inverted psum required by the binary subtraction (line \ref{mac2algo_minus_psum} of Algorithm \ref{mac2_algorithm}). 
The 6\textsuperscript{th} row stores the MAC2 result \textbf{P}. The 7\textsuperscript{th} row is a wide \textbf{Accumulator} to accumulate multiple MAC2 results that form a large dot product. 
        
The read and write operations of the dummy array are controlled by address and enable signals (blue signals in Fig. \ref{fig3_dummy_array_circuit}(a)) sent from the eFSM as described in Section \ref{bramac_mode_cim}. 
The access to 1\textsuperscript{st} -- 4\textsuperscript{th} rows during MAC2 is managed by both the decoder logic and a 2-to-4 demux. The 2-bit selection signal of the demux comes from the current two processing bits of the two inputs \textbf{I\textsubscript{1}} and \textbf{I\textsubscript{2}}, respectively. This allows calculating psum (line \ref{mac2algo_psum} of Algorithm \ref{mac2_algorithm}) using a look-up table \cite{lut_cim_tsmc}.
If \textbf{\{I\textsubscript{2}[i], I\textsubscript{1}[i]\}} is 2'b00, then the 1\textsuperscript{st} zero row will be read out and added to the 6\textsuperscript{th} row \textbf{P}. If \textbf{\{I\textsubscript{2}[i], I\textsubscript{1}[i]\}} is 2'b11, then the 4\textsuperscript{th} row \textbf{(W\textsubscript{1} + W\textsubscript{2})} will be read out and added to \textbf{P}. If \textbf{\{I\textsubscript{2}[i], I\textsubscript{1}[i]\}} is 2'b01 or 2'b10, then then the 2\textsuperscript{nd} row \textbf{W\textsubscript{1}} or the 3\textsuperscript{rd} row \textbf{W\textsubscript{2}} will be read out and added to \textbf{P}.

Since the dummy array copies data from the main BRAM array for computation, a coherency issue may arise where the main BRAM is being updated while the dummy array is still computing using the stale data. We leave it for the programmer/compiler to explicitly ensure the memory coherency similar to the explicit handling of the read-during-write behavior of Intel's BRAM \cite{M20K_Fmax}.

\subsubsection{Configurable Sign-Extension Mux} \label{config_sign_ext_mux}
Although not reflected in Algorithm \ref{mac2_algorithm}, the \textbf{W\textsubscript{1}} and \textbf{W\textsubscript{2}} vectors from the main BRAM need to be sign-extended before being copied to the dummy array in order to prevent overflow during MAC2. 
To support this, two configurable sign-extension muxes are added between the main BRAM and the dummy array. 
Each mux has five identical blocks, one of which is shown in Fig. \ref{fig3_dummy_array_circuit}(b). 
Since the main BRAM has a data width of 40 bits, it can copy five 8-bit, ten 4-bit, or twenty 2-bit elements to the dummy array simultaneously. 
Each of the five identical mux blocks can sign-extend one 8-bit element to one 32-bit element (blue crosses in Fig. \ref{fig3_dummy_array_circuit}(b)), or two 4-bit elements to two 16-bit elements (green crosses in Fig. \ref{fig3_dummy_array_circuit}(b)), or four 2-bit elements to four 8-bit elements (red crosses in Fig. \ref{fig3_dummy_array_circuit}(b)).  
Moreover, since a 2/4/8-bit MAC2 only requires a maximum bit-width of 5/9/17 bits to store the result, the proposed sign-extension mux can provide a higher bit-width required by MAC2.
This allows multiple sequential MAC2 results to be accumulated by adding the 6\textsuperscript{th} row (that stores the MAC2 result \textbf{P}) and the 7\textsuperscript{th} row (that stores the \textbf{Accumulator}) of the dummy array.

\subsubsection{Bit-Parallel SIMD Adder with Read/Write Circuits} \label{bit_parallel_adder}
The 160-bit SIMD adder in BRAMAC is designed using the conventional 1-bit full adder as shown in Fig. \ref{fig3_dummy_array_circuit}(c). 
It supports bit-parallel SIMD addition by configuring itself to twenty 8-bit adders, ten 16-bit adders, and five 32-bit adders for 2-bit, 4-bit, and 8-bit MAC2, respectively, giving a worst-case delay equal to 32-bit addition. 
The two operands \textbf{A} and \textbf{B} of the SIMD adder come from two sense amplifiers, \textbf{SA1} and \textbf{SA2} that compare the voltage differential of two bit-line pairs, (\textbf{BLA}, \textbf{BLbA}) and (\textbf{BLB}, \textbf{BLbB}). 
To support the addition followed by 1-bit shift-left operation (required in lines \ref{mac2algo_shift1_1} and \ref{mac2algo_shift1_2} of Algorithm \ref{mac2_algorithm}), a write-back mux \textbf{M1} before the write driver \textbf{WD1} is used to select either sum \textbf{S} from the current full adder or sum from the right full adder \textbf{S\textsubscript{Right}}. \textbf{M1} can also select \textbf{ramA} to copy the first data \textbf{W\textsubscript{1}} from the main BRAM. Similarly, a write-back mux \textbf{M2} before the write driver \textbf{WD2} is used to select between three signals: \textbf{B-bar} to perform inverting, \textbf{ramB} to copy the second data \textbf{W\textsubscript{2}} from the main BRAM, and 1'b0 to initialize either \textbf{P} (line \ref{mac2algo_init}) or the \textbf{Accumulator}. Both \textbf{M1} and \textbf{M2} are controlled by the eFSM.

\section{BRAMAC Variants} \label{BRAMAC_Optimization}

\subsection{BRAMAC with Two Synchronous Dummy Arrays (2SA)} \label{bramac_2sa}
This variant, called BRAMAC-2SA, has two synchronous dummy arrays that share the same clock domain as the main BRAM.
In this architecture, each dummy array is fed by one port of the main BRAM during weight copy. 
Since BRAMAC intrinsically supports multiplying the same input with many weights as discussed in Section \ref{mac2_dataflow}, this variant adopts an input-sharing approach to balance the data reuse. Specifically, in each MAC2 iteration, the two dummy arrays copy the same weights but process different inputs. The first dummy array receives two inputs $I_{1}, I_{2}$ and calculates $W_{1}I_{1} + W_{2}I_{2}$, while the second dummy array receives another two inputs $I_{3}, I_{4}$ and calculates $W_{1}I_{3} + W_{2}I_{4}$. 
    
\begin{figure}
    \centering
    \includegraphics[width=1\linewidth]{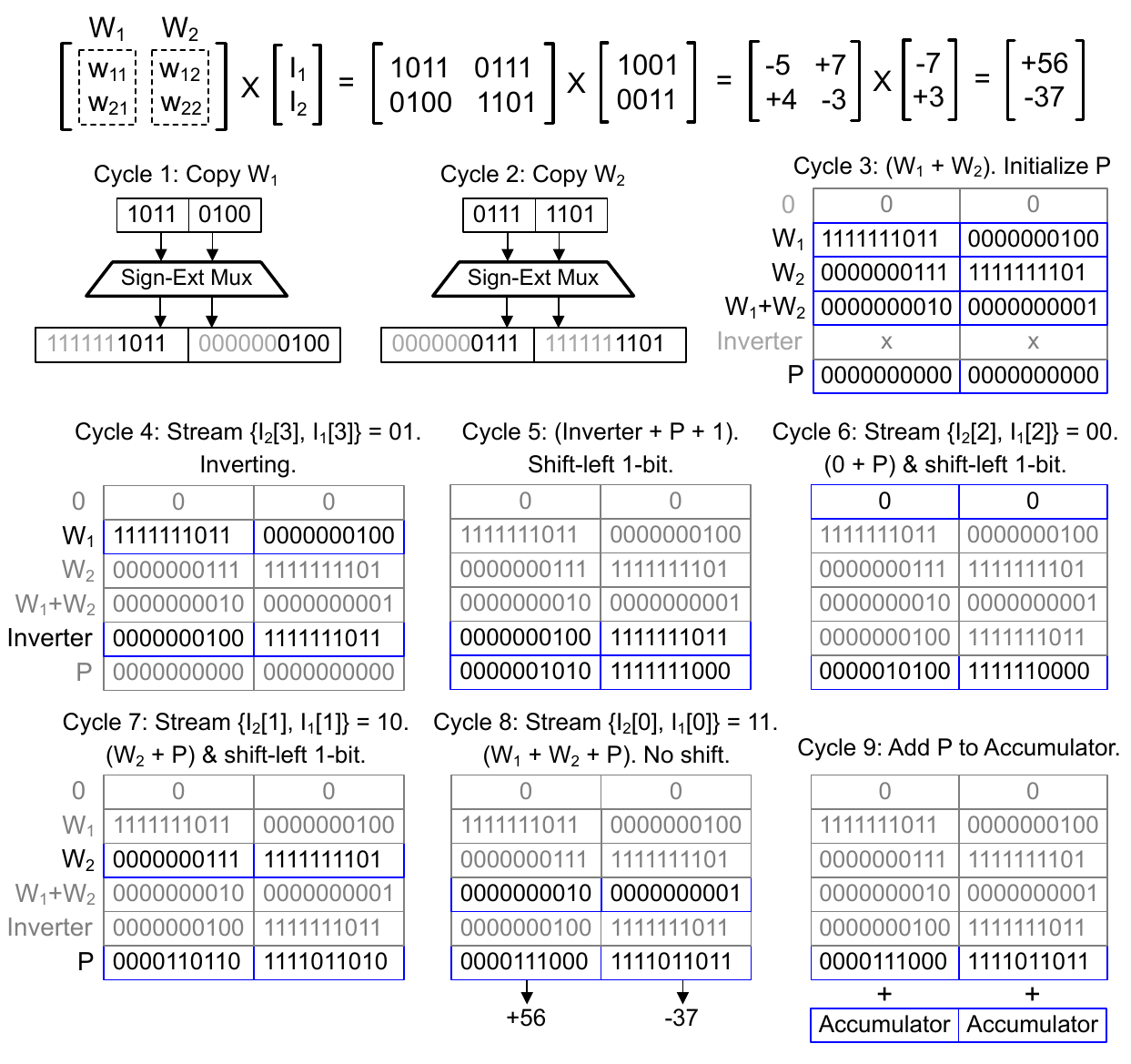}
    \caption{Example operation of one dummy array in BRAMAC-2SA for 4-bit MAC2. We are displaying 2 out of 10 lanes with 10-bit sign extension instead of 16 bits (due to space limitation).} 
    \label{fig4_dummy_array_operation}
\end{figure}

An example 4-bit MAC2 operation for one dummy array of BRAMAC-2SA is illustrated in Fig. \ref{fig4_dummy_array_operation}. Note that we are displaying 2 out of 10 lanes with 10-bit sign-extension due to space limitation (instead of 16-bit sign-extension as described in Section \ref{config_sign_ext_mux}). In \textbf{Cycle 1} and \textbf{Cycle 2}, \textbf{W\textsubscript{1}} and \textbf{W\textsubscript{2}} are sign-extended and copied to the dummy array. During these two cycles, the two inputs for each dummy array are also sent to BRAMAC-2SA through the CIM instruction and latched for further processing. In \textbf{Cycle 3}, \textbf{W\textsubscript{1}} and \textbf{W\textsubscript{2}} are read out and added. The sum is written back to the 4\textsuperscript{th} row to store \textbf{(W\textsubscript{1} + W\textsubscript{2})}. Simultaneously, the 6\textsuperscript{th} row \textbf{P} can also be initialized to zero.  In \textbf{Cycle 4}, the MSB of two inputs is streamed to the dummy array. The selected row \textbf{W1} is inverted to prepare for the binary subtraction. In \textbf{Cycle 5}, \textbf{Inverter} is added to \textbf{P}. The sum is shifted left by 1-bit and written back to \textbf{P}. The input streaming continues to \textbf{Cycle 8} where the LSB of two inputs is processed and the correct MAC2 result \textbf{P} is obtained. In \textbf{Cycle 9}, \textbf{P} is added to the 7\textsuperscript{th} \textbf{Accumulator} row. Then it can be initialized for the subsequent MAC2. 

The above example indicates that BRAMAC-2SA can complete a 4-bit MAC2 using 9 cycles. However, during the write-back phase of the last two cycles, i.e., \textbf{Cycle 8} and \textbf{Cycle 9}, the current two weights \textbf{W\textsubscript{1}} and \textbf{W\textsubscript{2}} are no longer needed in the dummy array since the current MAC2 result \textbf{P} is already obtained at the bit-parallel adder's output. As a result, these two cycles can also be used to copy the next two weights \textbf{W\textsubscript{3}} and \textbf{W\textsubscript{4}}, respectively as illustrated in Fig. \ref{fig5_dummy_array_pipeline}(a). Therefore, the 4-bit MAC2 in BRAMAC-2SA only requires 7 cycles to complete. This pipelining can also be applied to 2-bit and 8-bit MAC2. The only difference between 2-bit, 4-bit, and 8-bit MAC2 is the number of cycles spent for processing every input bit as described in line \ref{mac2algo_beginfor}-\ref{mac2algo_endfor} of Algorithm \ref{mac2_algorithm}. Thus, 2-bit and 8-bit MAC2 can take 5 and 11 cycles to complete, respectively.

\subsection{BRAMAC with One Double-Pumped Dummy Array (1DA)}
This variant, called BRAMAC-1DA, has only one dummy array to reduce the area overhead. 
Using one dummy array degrades the MAC throughput by 2$\times$ compared to BRAMAC-2SA, however, we propose to double-pump the dummy array with a 2$\times$ main BRAM clock frequency. 
Memory multi-pumping is a commonly used technique in FPGA design to improve the system throughput \cite{MultiPumpCache, FTDL}. 
The double-pumped dummy array doesn't add any additional area overhead compared to a synchronous dummy array. 
Rather, it only requires a separate clock routing during compilation. 
      
\begin{figure}
    \centering
    \includegraphics[width=1\linewidth]{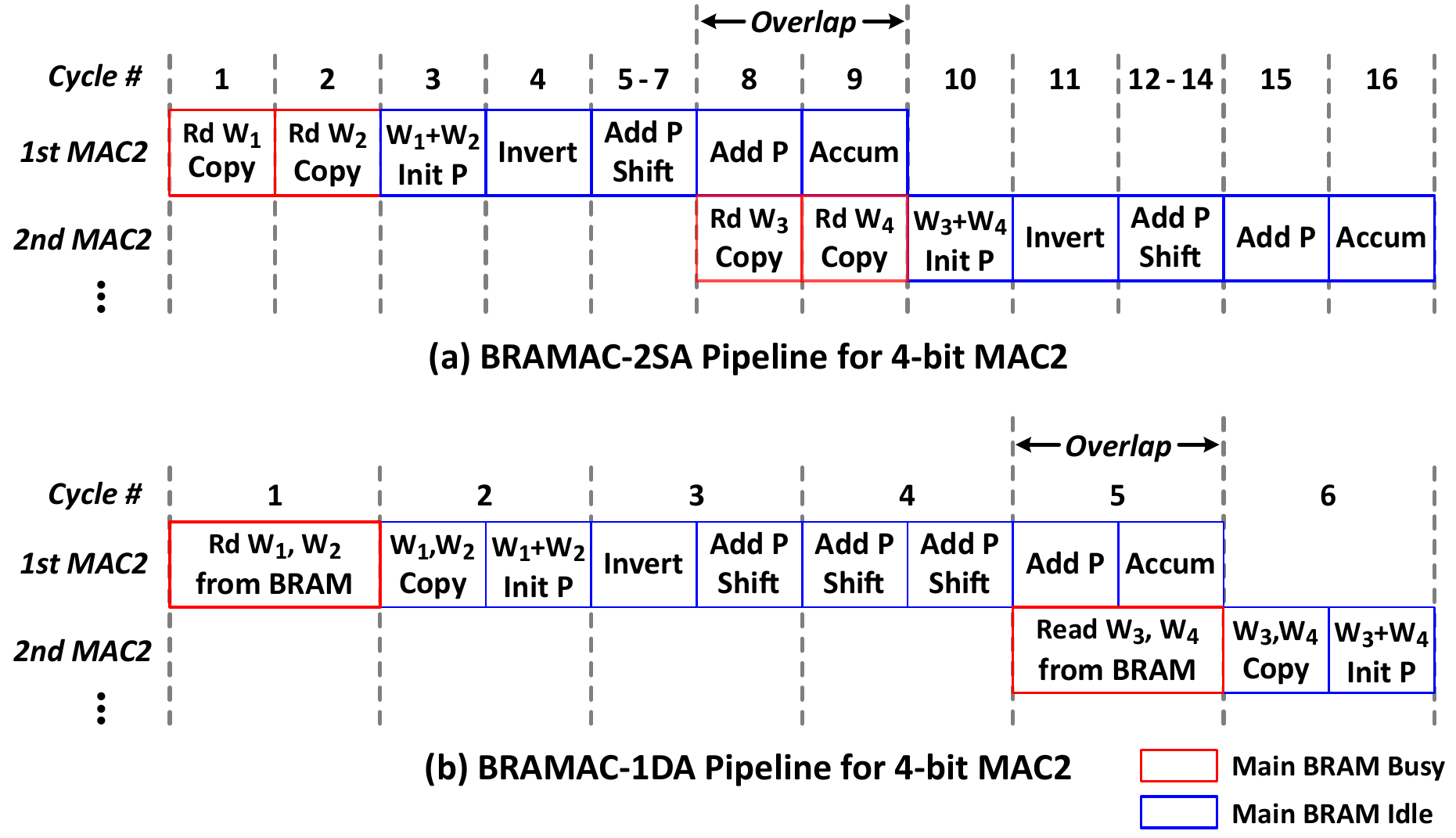}
    \caption{Pipeline diagram of 4-bit MAC2 in (a) BRAMAC-2SA and (b) BRAMAC-1DA.}
    \label{fig5_dummy_array_pipeline}
\end{figure}

Because the main BRAM and the dummy array only interact during weight copy, synchronization between them can be easily handled. Fig. \ref{fig5_dummy_array_pipeline}(b) shows the pipeline diagram of 4-bit MAC2 for BRAMAC-1DA. In \textbf{Cycle 1}, the main BRAM reads out two weights \textbf{W\textsubscript{1}} and \textbf{W\textsubscript{2}}. In the first half of \textbf{Cycle 2}, the dummy array copies \textbf{W\textsubscript{1}} and \textbf{W\textsubscript{2}} using its two write ports. Then the dummy array can compute the MAC2 using the same operation flow as BRAMAC-2SA, except that every cycle in BRAMAC-2SA is now half a cycle in BRAMAC-1DA. Similar to the pipelining optimization for BRAMAC-2SA, the main BRAM can start to read the next two weights \textbf{W\textsubscript{3}} and \textbf{W\textsubscript{4}} in \textbf{Cycle 5} while the dummy array is computing. As a result, the 4-bit MAC2 can be completed using 4 cycles. This pipelining can also be applied to 2-bit and 8-bit MAC2. 
Hence, 2-bit and 8-bit MAC2 can take 3 and 6 cycles to complete, respectively.

\subsection{Embedded FSM to Free Up BRAM Ports} \label{efsm}
Since the dummy array's behavior is deterministic for computing MAC2, we propose to control it using an eFSM. 
This eFSM receives a CIM instruction to trigger the MAC2 computation and control the dummy array's read/write access. The CIM instruction is only required when the main BRAM needs to send data to the dummy array (indicated by the red boxes in Fig. \ref{fig5_dummy_array_pipeline}). As a result, the main BRAM is busy for 2 cycles in BRAMAC-2SA and 1 cycle in BRAMAC-1DA. 
When the main BRAM is idle, it can perform normal read operations to feed LBs/DSPs or write operations to load the next tile of weights from off-chip DRAM, allowing tiling-based DNN acceleration. 
This is different from CCB and CoMeFa whose BRAM ports are always busy during CIM.

\begin{figure}
    \centering
    \includegraphics[width=1\linewidth]{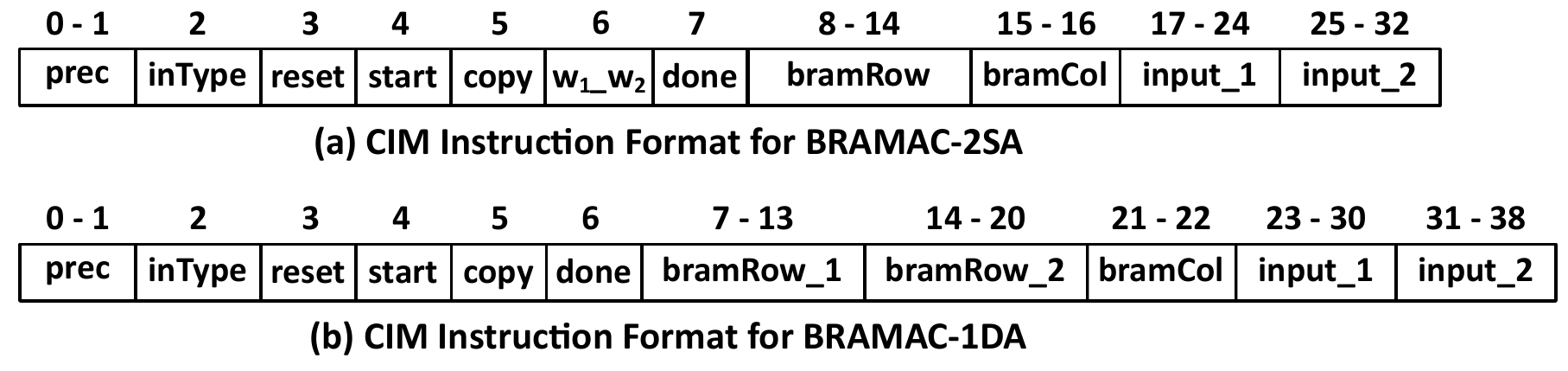}
    \caption{CIM instruction format for (a) BRAMAC-2SA and (b) BRAMAC-1DA.}
    \label{fig6_dummy_array_instruction}
\end{figure}

Fig. \ref{fig6_dummy_array_instruction}(a) and (b) show the proposed CIM instruction format for BRAMAC-2SA and BRAMAC-1DA, respectively. For BRAMAC-2SA, \textbf{bramRow} and \textbf{bramCol} are combined to form one BRAM address during each copy operation. On the other hand, BRAMAC-1DA needs to receive two BRAM addresses at the same time. This is achieved by using two BRAM row addresses \textbf{bramRow\-1} and \textbf{bramRow\-2} with a shared column address \textbf{bramCol}.

The two BRAMAC variants share some common control signals. The 2-bit \textbf{prec} specifies one of the three supported MAC2 precisions. The \textbf{inType} is used to indicate whether the two inputs are signed or unsigned. If the inputs are unsigned, then the inverting cycle can be skipped to improve performance. The \textbf{reset} resets the dummy array to the initial state and writes zero to its accumulator. When the \textbf{start} is enabled, BRAMAC is triggered to perform MAC2. 
The \textbf{copy} tells BRAMAC to copy the data read from the main BRAM to the dummy array, and an additional \textbf{w\textsubscript{1}\_w\textsubscript{2}} signal is needed for BRAMAC-2SA to indicate the currently copied data is $W_{1}$ or $W_{2}$. 
These two signals also allow the efficient pipelining optimization in Fig. \ref{fig5_dummy_array_pipeline} where the weight copy of the next MAC2 can be overlapped with computing the current MAC2.
The \textbf{done} indicates whether to read out the dummy array's accumulator. When it's enabled, the \textbf{bramCol} is used to select 40-bit data from the dummy array's accumulator row every cycle. As a result, between every two dot products, the main BRAM needs to be busy for 8 and 4 cycles to read out the accumulator in BRAMAC-2SA and BRAMAC-1DA, respectively. However, as the dummy array's accumulator has a size of 8/16/32-bit for 2/4/8-bit MAC precisions, it can process a maximum dot product size of 16/256/2048 before being read out to amortize this cost.

\section{Circuit-Level Evaluation} \label{circuit_evaluation}
\subsection{Tools and Baseline FPGA} \label{baseline_fpga}
We use COFFE \cite{COFFE}, an automatic FPGA transistor sizing tool, to model and optimize the area and delay of all BRAMAC components except for the eFSM which is implemented in SystemVerilog to verify its functionality. 
We use Synopsys Design Compiler with TSMC 28-nm technology to synthesize and get the area of the eFSM, which are 137 $\mu$m\textsuperscript{2} and 81 $\mu$m\textsuperscript{2} for BRAMAC-2SA and BRAMAC-1DA, respectively after scaling to 22-nm. 
We get the area of an M20K block from COFFE by interpolating between 16~kb and 32~kb BRAMs. For delay estimation, COFFE runs Hspice simulations using the 22~nm Predictive Technology Model \cite{PTM}.

\begin{table}
  \centering
  \caption{Resource Counts and Area Ratio of the Baseline Arria 10 GX900 FPGA.}
  \renewcommand{\arraystretch}{1.3}
  \begin{tabular}{ | m{2.5cm}<{\centering} | m{1.4cm}<{\centering} | m{1.6cm}<{\centering} | } 
      \hline 
      \rowcolor{light-gray} 
      \textbf{Resource} & \textbf{Count} & \textbf{Area Ratio} \\ 
      \hline
      Logic Blocks (LBs) & 33920 & 70.4\% \\
      \hline
      DSP Units  & 1518  & 9.5\%  \\
      \hline
      BRAMs (M20K)  & 33920 & 20.1\% \\
      \hline
  \end{tabular}
  
  \label{Arria10_GX900}
\end{table}

For the baseline FPGA in the remainder of this paper, we use an Arria-10 GX900 device \cite{Arria10} at the fastest speed grade (10AX090H1F34E1SG) whose resource information is shown in Table \ref{Arria10_GX900}. The Arria-10 device family is fabricated using 20-nm technology similar to COFFE's simulation setup. The area ratio for each resource type is estimated based on the area model in \cite{TILT}. The proposed BRAMAC architecture enhances the baseline FPGA by replacing all M20K blocks with either BRAMAC-2SA or BRAMAC-1DA.

\begin{table*}
  \centering
  \caption{Key Features of BRAMAC and Prior State-of-the-art MAC Architectures for FPGA}
  \renewcommand{\arraystretch}{1.2}

  \begin{threeparttable}
      
      \begin{tabularx}{0.9\textwidth}{|L|L|S||M|M|L|L|L|L|L|} 
          \hline 
          \rowcolor{light-gray} 
          
          \mcccll{\cellcolor{light-gray}{Architecture}} & 
          \thead{eDSP\\ \cite{EmbracingDiversity}} & 
          \thead{PIR-DSP\\ \cite{PIR_DSP}} & 
          \thead{CCB\\ \cite{CCB}} & 
          \thead{CoMeFa-D\\ \cite{CoMeFa}} & 
          \thead{CoMeFa-A\\ \cite{CoMeFa}} & 
          \thead{BRAMAC-\\ 2SA} &
          \thead{BRAMAC-\\ 1DA} \\ 
          \hline
          \hline 
        
          \mcccll{\thead{Modified FPGA Block}} & 
            \thead{DSP} & 
            \thead{DSP} &
            \thead{BRAM} & 
            \thead{BRAM} &
            \thead{BRAM} &
            \thead{BRAM} &
            \thead{BRAM} \\
          \hline
          
          \mcccll{\thead{Supported MAC Precision (-bit)}} & 
            \thead{4, 8} & 
            \thead{2, 4, 8} & 
            \thead{Arbitrary} & 
            \thead{Arbitrary} &
            \thead{Arbitrary} &
            \thead{2, 4, 8} & 
            \thead{2, 4, 8} \\
          \hline
          
          \mcccll{\thead{Area Overhead (Block)}} & 
            \thead{12\%} & 
            \thead{28\%} &
            \thead{16.8\%} & 
            \thead{25.4\%} & 
            \thead{8.1\%} & 
            \thead{33.8\%} & 
            \thead{16.9\%} \\ 
          \hline
          
          \mcccll{\thead{Area Overhead (Core)}} & 
            \thead{1.1\%} & 
            \thead{2.7\%} &
            \thead{3.4\%} & 
            \thead{5.1\%} & 
            \thead{1.6\%} & 
            \thead{6.8\%} & 
            \thead{3.4\%} \\ 
          \hline
        
          \mcccll{\thead{Clock Period Overhead \\ over the Baseline FPGA Block}} & 
            \thead{0\%} & 
            \thead{30\%} &
            \thead{60\%} & 
            \thead{25\%} & 
            \thead{150\%} & 
            \thead{10\%} & 
            \thead{46\%} \\ 
          \hline
          
          \mccl{\multirow{3}{*}{\thead{\# of MACs in Parallel / \\ MAC Latency (Cycles) \tnote{1}}}} & \thead{2-bit} & 
            \thead{8 / 1} & 
            \thead{24 / 1} &
            \thead{160 / 16} & 
            \thead{160 / 16} & 
            \thead{160 / 16} & 
            \thead{80 / 5} & 
            \thead{40 / 3} \\ 
          \cline{3-10}
        
          \mccl{\thead{}} & \thead{4-bit} & 
            \thead{8 / 1} & 
            \thead{12 / 1} &
            \thead{160 / 42} & 
            \thead{160 / 42} & 
            \thead{160 / 42} & 
            \thead{40 / 7} & 
            \thead{20 / 4} \\ 
          \cline{3-10}
        
          \mccl{\thead{}} & \thead{8-bit} & 
            \thead{4 / 1} & 
            \thead{6 / 1} &
            \thead{160 / 113} & 
            \thead{160 / 113} & 
            \thead{160 / 113} & 
            \thead{20 / 11} & 
            \thead{10 / 6} \\ 
          \hline
          
          \mcccll{\thead{Design Complexity}} & 
            \thead{Very Low} & 
            \thead{Very Low} &
            \thead{High} & 
            \thead{Low} & 
            \thead{Medium} & 
            \thead{Low} & 
            \thead{Medium} \\ 
          \hline
      
      \end{tabularx}
      
      \begin{tablenotes}
        \item[1] For DSP architectures, the accumulator size for each MAC precision is the same as that in the baseline DSP. \\ 
        For BRAM architectures, the accumulator sizes for 2-bit, 4-bit, and 8-bit MACs are 8-bit, 16-bit, and 27-bit, respectively. The MAC latency is reported based on unsigned multiplication for CCB and CoMeFa, and 2's complement multiplication for BRAMAC.
      \end{tablenotes}
      
  \end{threeparttable}
  
    \label{EnhancedFPGA}
\end{table*}

\subsection{Design Choice for Adder}
As the SIMD adder in BRAMAC has a worst-case delay of 32-bit addition during 8-bit MAC2, a ripple-carry adder (RCA) can significantly increase the critical path delay of the dummy array and become the frequency bottleneck of BRAMAC. Hence, we also explore two variants of fast adders \cite{fastadder}: Carry Lookahead Adder (CLA) with a 4-bit carry lookahead generator using mirror implementation, and Carry Bypass Adder (CBA) with 4-bit Manchester carry chain using dynamic logic. We use COFFE to automatically size the carry-out generator, the carry lookahead generator, and the Manchester carry chain to obtain the best area-delay trade-off for RCA, CLA, and CBA, respectively. 
    
\begin{figure}
    \centering
    \includegraphics[width=1\linewidth]{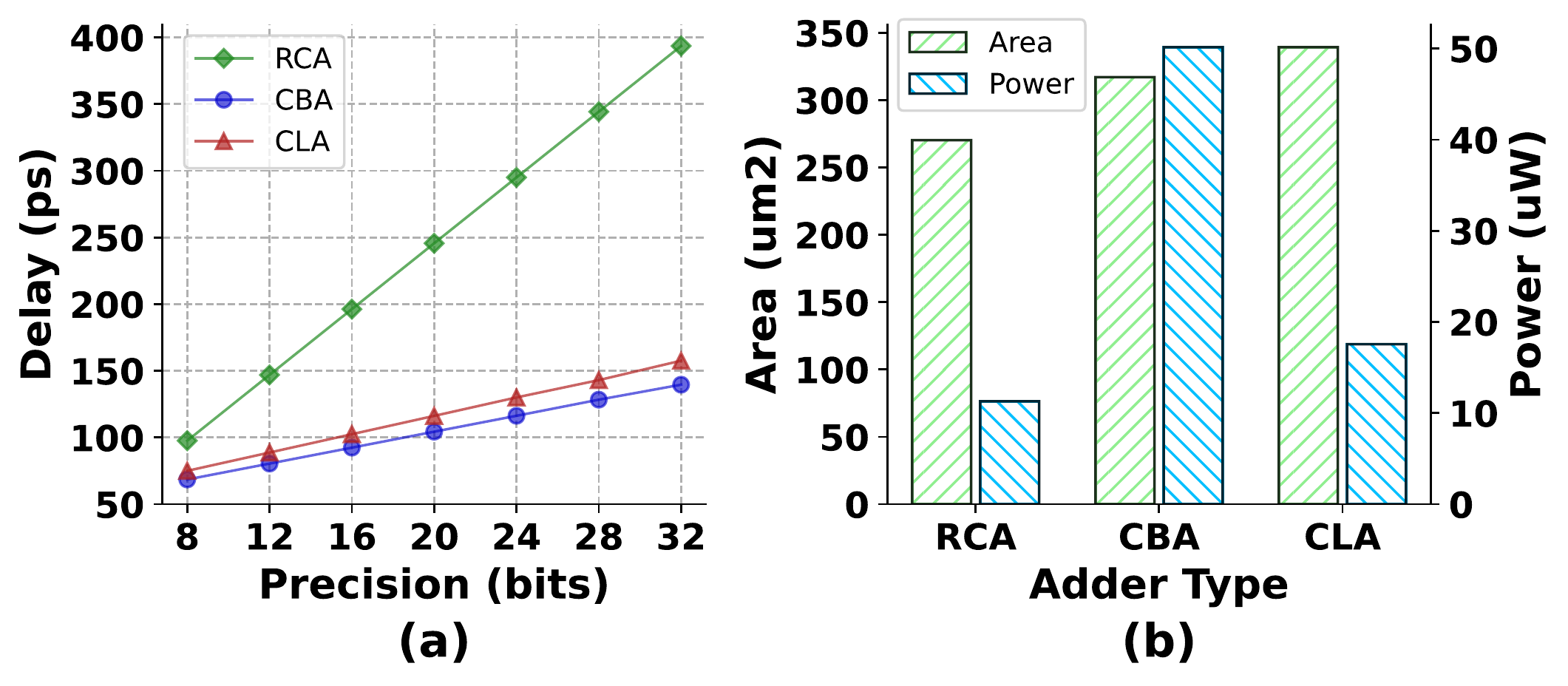}
    \caption{Comparison between RCA, CBA, and CLA: (a) Delays vs. precision. (b) Area and power at 32-bit precision.}
    \label{adders_delay_area_power}
\end{figure}

Fig. \ref{adders_delay_area_power} illustrates the performance, area, and power of three different adders RCA, CBA, and CLA based on COFFE simulations. As shown in Fig. \ref{adders_delay_area_power}(a), the performance gap between RCA and two other fast adders CBA/CLA becomes larger as the adder precision increases. At the highest adder precision, i.e., 32-bit accumulation during 8-bit MAC, RCA has a delay of 393.6 ps, which is 2.8$\times$ slower than CBA (139.6 ps) and 2.5$\times$ slower than CLA (157.6 ps). As illustrated in Fig. \ref{adders_delay_area_power}(b), all three adders have similar areas, but CBA has the highest power consumption of 50.2 $\mu$W, which is 4.44$\times$ and 2.86$\times$ higher than RCA (11.3 $\mu$W) and CLA (17.6 $\mu$W), respectively. This is because that CBA employs the dynamic Manchester carry chain which is faster but more power-hungry than static CMOS logic. Overall, CLA has the best trade-off between delay, area, and power. Hence, we adopt CLA in BRAMAC for the remainder of our evaluation.
    
\subsection{BRAMAC Area and Frequency} \label{bramac_area_frequency}
\begin{figure}
    \centering
    \includegraphics[width=1\linewidth]{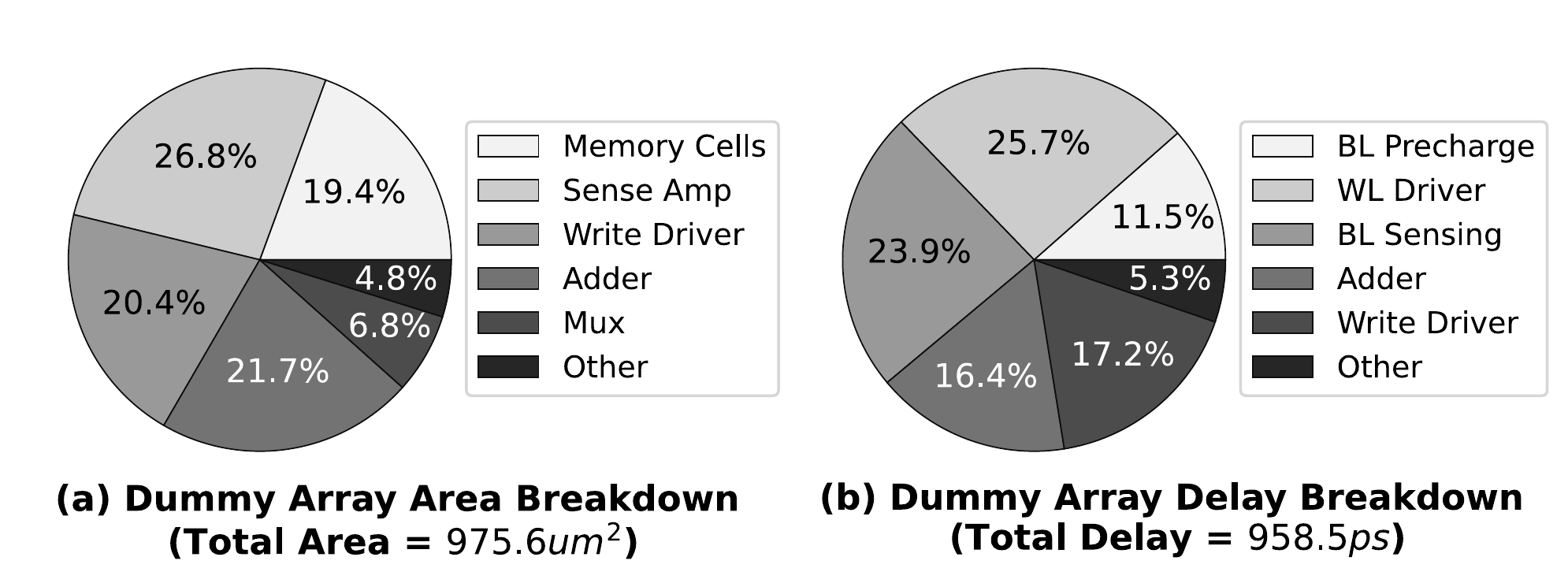}
    \caption{(a) Area and (b) delay breakdown of the dummy array.}
    \label{area_delay_breakdown}
\end{figure}

Fig. \ref{area_delay_breakdown}(a) illustrates the area breakdown of BRAMAC's dummy array. The total area of a dummy array is 975.6 $\mu$m\textsuperscript{2}, which represents an area increase of 16.9\% compared to the baseline M20K. Since M20K constitutes 20.1\% area of the baseline FPGA, this area overhead is equivalent to only 3.4\% increase in the FPGA core area. 
Note that we ignore the area overhead of eFSM in our later evaluation because COFFE's area model doesn't include any BRAM control logic and some M20K components such as error correction circuits \cite{M20K}. Given that the eFSMs of BRAMAC-2SA/BRAMAC-1DA are equivalent to only 1.4\%/2.4\% of the baseline M20K area, it's expected that the area overhead of BRAMAC doesn't change compared to the baseline M20K when a more accurate area model is adopted.

Fig. \ref{area_delay_breakdown}(b) shows the critical path delay breakdown of BRAMAC's dummy array. With only 7 rows, the dummy array's bitline parasitic load is significantly reduced compared to the main BRAM. 
As a result, it can precharge and discharge much faster, giving less than 1 ns critical path delay. 
This suggests that the dummy array itself is able to run at a maximum frequency (F\textsubscript{max}) of 1 GHz independent from M20K whose F\textsubscript{max} is 730 MHz in Arria-10 \cite{M20K_Fmax}. 
For BRAMAC-1DA, this limits the F\textsubscript{max} of M20K to 500 MHz in CIM mode. While this is less than the typical BRAM F\textsubscript{max}, realistic FPGA delays are usually constrained by soft logic and routing, and it is unlikely that a design on Arria-10 will achieve a frequency higher than 500 MHz. 
For BRAMAC-2SA, the critical path occurs during the weight copy where the write-back phase can only start after reading out data from the main BRAM. 
Hence, the F\textsubscript{max} of BRAMAC-2SA is dependent on M20K. Specifically, the dummy array's write driver has a delay of 165 ps, leading to a 1.1$\times$ lower F\textsubscript{max} compared to the baseline M20K.

\subsection{Comparison with Other MAC Architectures on FPGA} \label{FPGABaseline}
We compare BRAMAC with other state-of-the-art architectures for MAC on FPGA, including eDSP \cite{EmbracingDiversity}, PIR-DSP \cite{PIR_DSP}, CCB \cite{CCB}, and CoMeFa \cite{CoMeFa}. 
All architectures use the same baseline Arria-10 FPGA as described in Section \ref{baseline_fpga}.
Each architecture replaces the corresponding FPGA block in the baseline with its proposed new block. The key features for each studied architecture are summarized in Table \ref{EnhancedFPGA}. 

Due to bit-serial arithmetic, CCB and CoMeFa have the highest flexibility in the supported precision. However, their proposed bit-serial algorithms for fixed-point multiplication only work for unsigned numbers, while eDSP, PIR-DSP, and BRAMAC can support 2's complement MAC. Although BRAMAC-2SA has the highest area overhead, it achieves the highest frequency compared to other BRAM architectures.
The two DSP architectures have the lowest design complexity as they can be implemented in digital CAD flow, while BRAM design typically involves analog components and manual layout effort \cite{COFFE}. Among all BRAM architectures, CCB has the highest design complexity as it needs an extra voltage supply. CoMeFa-A and BRAMAC-1DA have medium design complexity since they require novel timing design techniques---sense amplifier cycling and a double-pumped clock, respectively.

\section{Application-Level Evaluation} \label{application_evaluation}

\subsection{Peak MAC Throughput Comparison} \label{peak_mac_throughput_comparison}
\begin{figure*}
    \centering
    \includegraphics[width=1\linewidth]{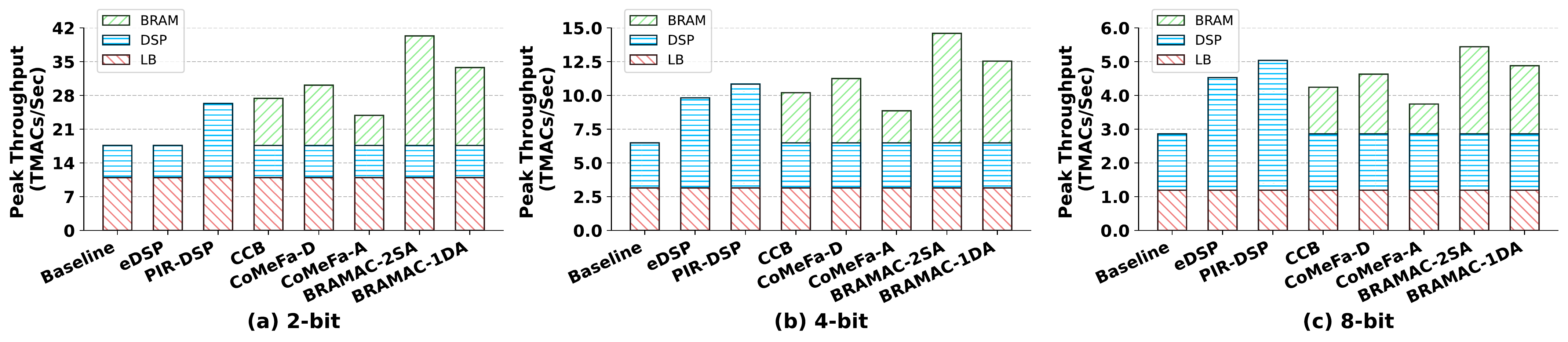}
    \caption{Peak MAC throughput of different architectures for various MAC precisions: (a) 2-bit, (b) 4-bit, (c) 8-bit.}
    \label{peak_mac_throughput}
\end{figure*}

We compare the peak MAC throughput of the baseline FPGA with those of enhanced FPGAs that employ BRAMAC and other MAC architectures studied in Section \ref{FPGABaseline}. We consider three MAC precisions: 2-bit multiply (with an 8-bit accumulator), 4-bit multiply (with a 16-bit accumulator), and 8-bit multiply (with a 27-bit accumulator). The peak MAC throughput of each resource type is determined as follows:

\begin{para_noindent}
    \textit{(1) LB:} We synthesize, place, and route one MAC unit using only LBs in Quartus to obtain its F\textsubscript{max} and resource utilization. We then follow the same methodology as \cite{CCB, CoMeFa} to calculate the total MAC throughput by optimistically assuming that all LBs can be used at the same F\textsubscript{max}. 
    
    \textit{(2) DSP:} The Arria-10 DSP has two 18$\times$19 multipliers, each can implement one 8-bit MAC, two 4-bit MACs, or four 2-bit MACs using DSP packing described in \cite{dspPacking}. We run Quartus to generate a DSP in m18x18\textunderscore sumof2 mode and find its F\textsubscript{max} to be 549 MHz. We use the same F\textsubscript{max} for eDSP but a 1.3$\times$ lower F\textsubscript{max} for PIR-DSP based on its reported F\textsubscript{max}. 
    
    \textit{(3) BRAM:} We use Quartus to generate the baseline M20K in simple dual-port mode and find its F\textsubscript{max} to be 645 MHz. BRAMAC-2SA and BRAMAC-1DA would run at 586 MHz (1.1$\times$ lower) and 500 MHz, respectively, while CCB, CoMeFa-D, and CoMeFa-A would run 1.6$\times$, 1.25$\times$, and 2.5$\times$ slower, respectively based on their reported F\textsubscript{max} degradation. 
\end{para_noindent}

Fig. \ref{peak_mac_throughput} shows the peak MAC throughput breakdown in TeraMACs/sec for different architectures and MAC precisions. Compared to the baseline Arria-10 device, BRAMAC-2SA/BRAMAC-1DA can improve the peak throughput by 2.6$\times$/2.1$\times$, 2.3$\times$/2.0$\times$, and 1.9$\times$/1.7$\times$ for 2-bit, 4-bit, and 8-bit MAC, respectively. 
Although CCB and CoMeFa can compute 160 MACs in parallel, they suffer from long-latency bit-serial arithmetic, leading to lower throughput than BRAMAC. Compared to low-precision DSP architectures, BRAMAC-2SA can deliver higher MAC throughput across all precisions, while BRAMAC-1DA's throughput is only slightly lower than PIR-DSP for 8-bit MAC. Note that BRAMAC is an enhanced BRAM architecture, therefore doesn't preclude the use of eDSP or PIR-DSP on the same FPGA. The combination of BRAMAC and eDSP/PIR-DSP can further boost an FPGA's MAC throughput.

\subsection{BRAM Utilization Efficiency for DNN Model Storage}
\begin{figure}
    \centering
    \includegraphics[width=1\linewidth]{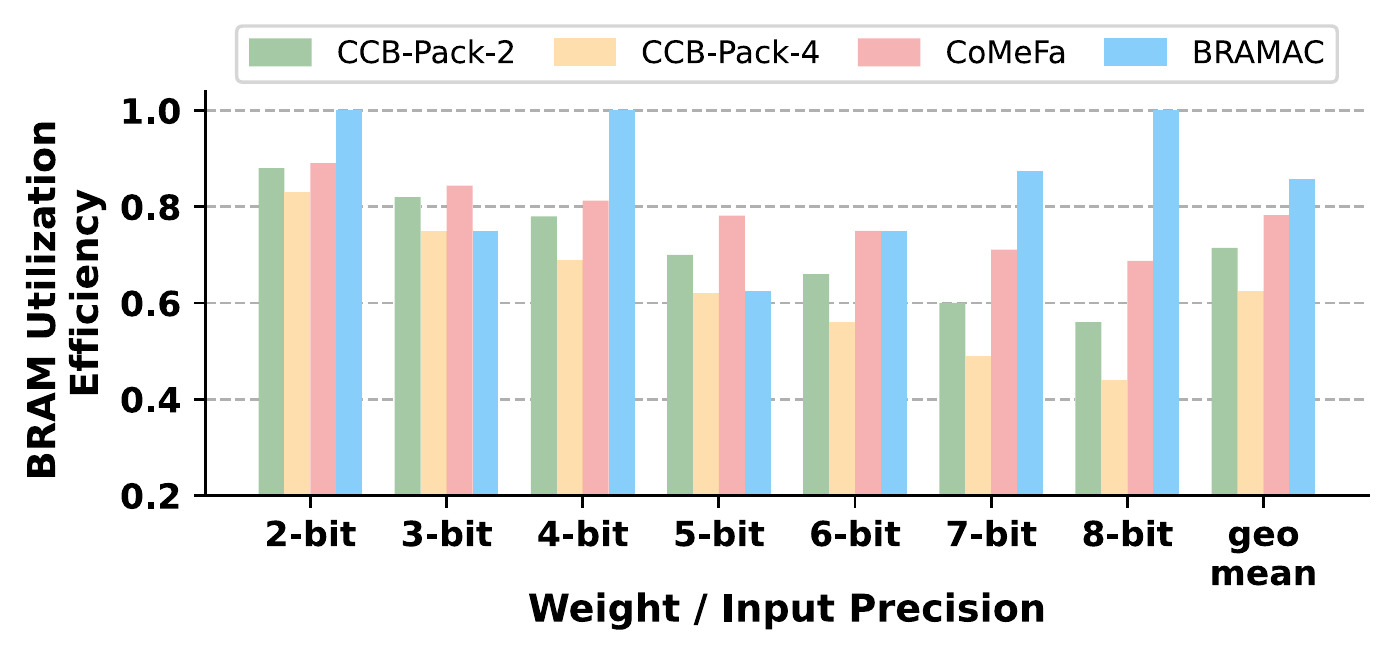}
    \caption{Comparison of BRAM utilization efficiency for DNN model storage at different precisions.}
    \label{bram_utilization}
\end{figure}

Since BRAMAC computes MAC in a separate dummy array that is fully decoupled from the main BRAM, it can store a DNN model efficiently. Fig. \ref{bram_utilization} compares the BRAM utilization efficiency between BRAMAC, CCB, and CoMeFa for storing DNN models with different precisions from 2- to 8-bit. Here, utilization efficiency is defined as the effective capacity ratio of a BRAM that can be used to store weight. A higher utilization efficiency can store the DNN model using fewer BRAM blocks, saving both area and power consumption. For CCB, we examine two variants, CCB-Pack-2 and CCB-Pack-4, that map 2 and 4 sequential bit-serial MACs to the same BRAM column, respectively.

BRAMAC can achieve 100\% utilization for 2-bit, 4-bit, and 8-bit precisions. 
Other precisions can be stored in BRAMAC with lower efficiency by sign-extending them to 4-bit or 8-bit. 
Despite this, BRAMAC still achieves the highest average BRAM utilization efficiency which is 1.3$\times$ and 1.1$\times$ better compared to CCB and CoMeFa, respectively. 
This is because CCB and CoMeFa use extra BRAM space to store temporary products and partial sums, while BRAMAC stores temporary results only in the dummy array. 
For CCB, a higher packing factor computes more sequential MACs before a slow in-memory reduction, giving a higher performance at the cost of more BRAM usage to save a copy of the input vector. 
On the other hand, CoMeFa offers a one-operand-outside-RAM mode that streams the input vector, avoiding a copy to BRAM which improves utilization efficiency when compared to CCB.

\subsection{Performance Improvement over CCB and CoMeFa}
We use general matrix-vector multiplication (GEMV) to benchmark and compare the application performance of BRAMAC, CCB, and CoMeFa. 
We choose BRAMAC-1DA for this experiment because it has a similar BRAM area and frequency overhead as CCB/CoMeFa. We assume that there is only one BRAM block available to perform the computation. 
This approach captures the performance of an architecture normalized to BRAM utilization. 
We consider both persistent and non-persistent (tiling-based) computations that exclude and include the cycles needed for loading the matrix data to the single BRAM block, respectively. 
Since the data mapping and computation flow of the three studied architectures are deterministic, we use a detailed analytical model to map a given GEMV workload to each architecture and count the number of cycles required. 
In addition to the latency of MAC, our analytical model accounts for latency associated with copying the input vector and reading out the accumulation results in each architecture. 

\begin{figure} 
    \centering
    \includegraphics[width=1\linewidth]{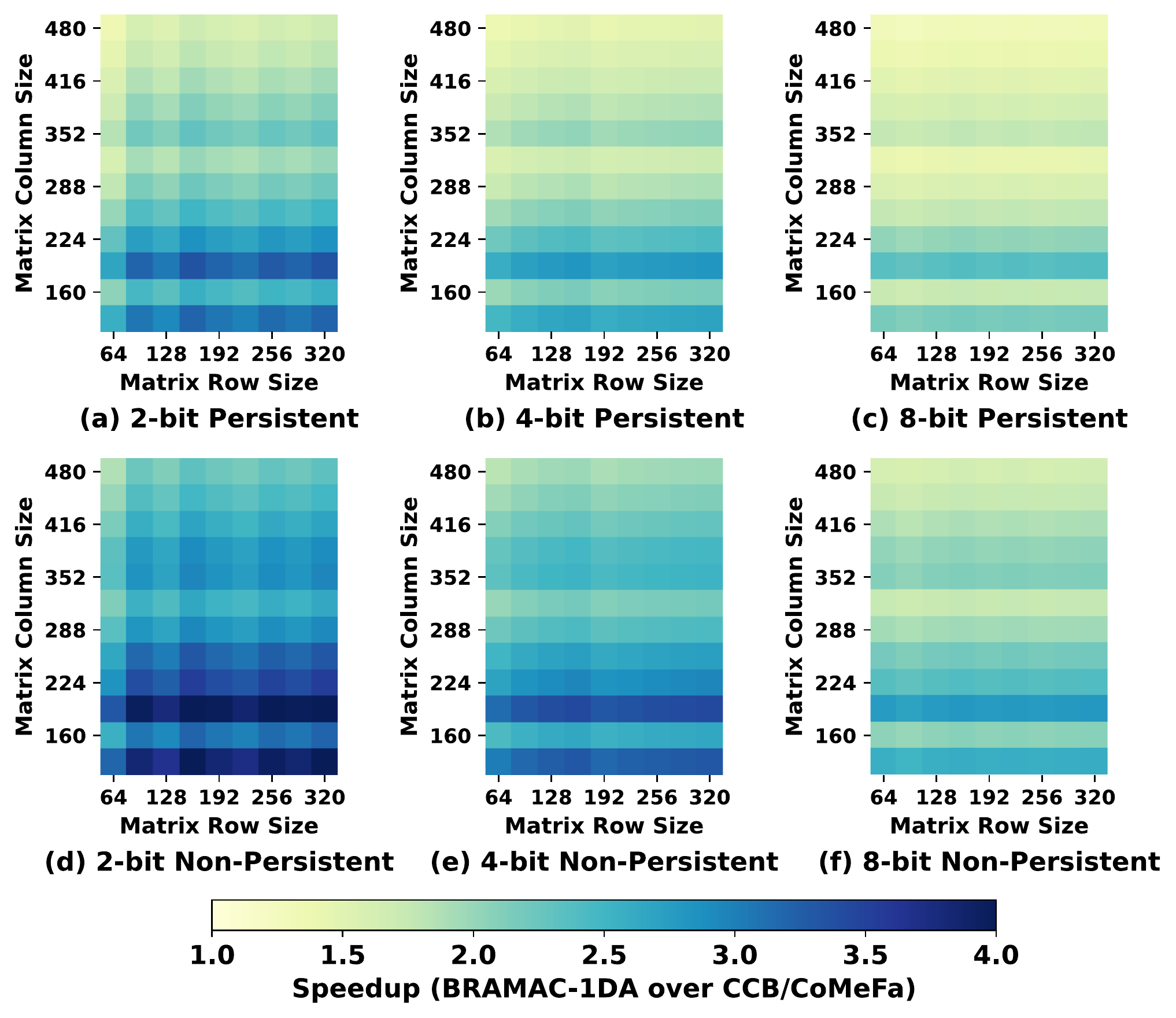}
    \caption{Speedup (based on cycles) of BRAMAC-1DA over CCB/CoMeFa for GEMV with different matrix sizes, precisions, and computation styles.}
    \label{gemv_speedup}
\end{figure}

Fig. \ref{gemv_speedup} illustrates the speedup of BRAMAC-1DA over CCB and CoMeFa when performing GEMV with different matrix sizes, precisions (2-bit, 4-bit, 8-bit), and computation styles (persistent and non-persistent). 
Overall, BRAMAC-1DA achieves up to 3.3$\times$/2.8$\times$/2.4$\times$ (and 4.1$\times$/3.4$\times$/2.8$\times$) speedups for 2/4/8-bit persistent (and non-persistent) GEMV. 

At the same precision, BRAMAC-1DA achieves higher speedup for non-persistent computation thanks to its eFSM that allows loading the next matrix tile while computing on the current tile.
Regarding different precisions, the speedup of BRAMAC-1DA decreases as the precision increases. This is because a higher precision directly reduces the computation parallelism of BRAMAC-1DA by 2$\times$, and it takes more cycles to process more input bits. On the other hand, CCB/CoMeFa only sacrifice latency but not parallelism at higher precision. Nevertheless, BRAMAC-1DA still achieves better performance for all cases due to its overall MAC throughput improvement over CCB/CoMeFa as discussed in Section \ref{peak_mac_throughput_comparison}. Note that CCB/CoMeFa's bit-serial algorithms for fixed-point multiplication only support unsigned numbers. 
It's expected that they require much higher latency when supporting 2's complement MAC.

Along the matrix row size, the speedup of BRAMAC-1DA is mainly affected by the vectorization efficiency, and this effect is more pronounced at a lower precision. 
For example, consider the 2-bit persistent case in Fig. \ref{gemv_speedup}(a), where BRAMAC-1DA can compute 20 outputs simultaneously. 
If the matrix row size is 64, i.e., the first column in Fig. \ref{gemv_speedup}(a), 
then at least 4 iterations are required to compute an output vector of size 64, with only $64/80 = 80\%$ useful computation in BRAMAC-1DA. 
On the other hand, if the matrix row size is 160, i.e., the fourth column in Fig. \ref{gemv_speedup}(a), then the output vector divides perfectly into 8 iterations at $100\%$ efficiency, thus giving better speedup as indicated by the darker color of the fourth column compared to the first column. Similar trends exist in 4-bit and 8-bit cases but are less pronounced. 

Along the matrix column size, the speedup of BRAMAC-1DA is determined by not only the vectorization efficiency but also the achievable packing factor of CCB/CoMeFa. 
For example, consider the 8-bit non-persistent case in Fig. \ref{gemv_speedup}(f). If the matrix column size is 480, i.e., the top row in Fig. \ref{gemv_speedup}(f), then CCB/CoMeFa can perform 3 sequential MACs on the same BRAM column before a slow in-memory reduction to amortize the reduction's latency cost. On the other hand, if the matrix column size is 128, i.e., the bottom row in Fig. \ref{gemv_speedup}(f), then a reduction is necessary for CCB/CoMeFa after every bit-serial MAC, resulting in much longer latency. On the contrary, BRAMAC's dummy array doesn't require a special reduction operation. Rather, it performs in-place accumulation at the end of every MAC2.

\begin{figure}
    \centering
    \includegraphics[width=1\linewidth]{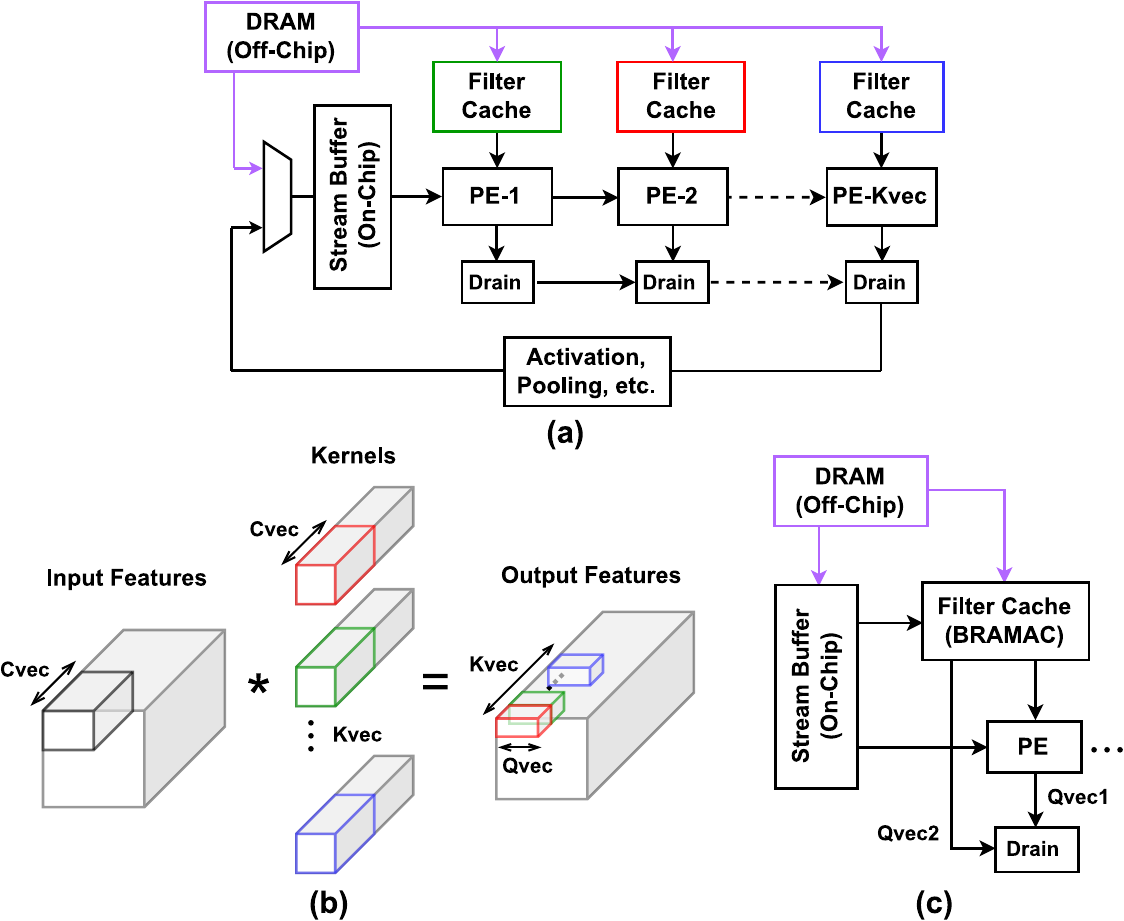}
    \caption{DLA's (a) architecture and (b) computation parallelism across different axes for CNNs. (c) The architecture of DLA-BRAMAC (with one PE shown).}
    \label{dla_bramac_architecture}
\end{figure}

\begin{table*}
  \centering
  \caption{Optimal Configurations of DLA and DLA-BRAMAC for AlexNet and ResNet-34}
  \renewcommand{\arraystretch}{1.2}
  
  \begin{threeparttable}
      \begin{tabularx}{0.9\textwidth}{|M|M|S||M S S|L S S|L S S|} 
      \hline 
      
      \rowcolor{light-gray}
      \mcccll{} & 
        \mcccl{DLA} & 
        \mcccl{DLA-BRAMAC-2SA} & 
        \mcccl{DLA-BRAMAC-1DA} \\ 
      \cline{4-12}
      
      \rowcolor{light-gray}
      \mcccll{\multirow{-2}{*}{ \diagbox[height=2.4\line]{\rlap{\enspace\raisebox{0.5ex}{Model}}}{\raisebox{-0.5ex}{Accelerator}} }} & 
        \thead{ Config\tnote{1}} & 
        \thead{ DSPs } &
        \thead{ BRAMs } &
        \thead{ Config\tnote{2}} & 
        \thead{ DSPs } &
        \thead{ BRAMs } &
        \thead{ Config\tnote{2}} & 
        \thead{ DSPs } &
        \thead{ BRAMs } \\ 
      \hline 
    
      \mccl{ \multirow{3}{*}{ \thead{ \ \ \ \ AlexNet \ \ \ \ }}}  & \thead{ 2-bit } & 
        \thead{ (2, 16,  96) } & 
        \thead{ 1152 } &
        \thead{ 352 } & 
        
        \thead{ (1+2, 24, 140) } & 
        \thead{ 1260 } & 
        \thead{ 1128 } & 
        
        \thead{ (2+2, 16, 100) } &
        \thead{ 1200 } &
        \thead{ 816 }  \\ 
        
      \mccl{\thead{}} & \thead{ 4-bit } & 
        \thead{ (3, 16,  32) } & 
        \thead{ 1152 } &
        \thead{ 544 } & 

        \thead{ (1+2, 16, 100) } & 
        \thead{ 1200 } & 
        \thead{ 1600 } & 

        \thead{ (1+1, 12, 130) } &
        \thead{ 1170 } &
        \thead{ 1080 } \\ 
    
      \mccl{\thead{}} & \thead{ 8-bit } &  
        \thead{ (3, 12, 24) } & 
        \thead{ 1296 } &
        \thead{ 868 } & 

        \thead{ (2+2, 10, 50) } & 
        \thead{ 1500 } & 
        \thead{ 1740 } & 

        \thead{ (1+1, 8, 100) } &
        \thead{ 1200 } &
        \thead{ 1664 }  \\ 
      \hline

      \mccl{ \multirow{3}{*}{ \thead{ \ ResNet-34\ } } } & \thead{ 2-bit } & 
        \thead{ (4, 12, 72) } & 
        \thead{ 1296 } &
        \thead{ 792 } & 

        \thead{ (1+2, 16, 140) } & 
        \thead{ 840 } & 
        \thead{ 832 } & 

        \thead{ (2+2, 22, 80) } &
        \thead{ 1320 } &
        \thead{ 924 }  \\ 
    
      \mccl{ \thead{} } & \thead{ 4-bit } & 
        \thead{ (3, 8, 64) } & 
        \thead{ 1152 } &
        \thead{ 736 } & 

        \thead{ (2+2, 12, 70) } & 
        \thead{ 1260 } & 
        \thead{ 972 } & 

        \thead{ (1+1, 16, 90) } &
        \thead{ 1080 } &
        \thead{ 1056 } \\ 
    
      \mccl{ \thead{} } & \thead{ 8-bit } &  
        \thead{ (3, 4, 64) } & 
        \thead{ 1152 } &
        \thead{ 1452 } & 
        
        \thead{ (2+2, 6, 65) } & 
        \thead{ 1170 } & 
        \thead{ 1530 } & 

        \thead{ (1+1, 12, 65) } &
        \thead{ 1170 } &
        \thead{ 1788 }  \\ 
      \hline
      
    \end{tabularx}
  
    \begin{tablenotes}
      \item[1] The configuration value for DLA has the form of (Qvec, Cvec, Kvec).
      \item[2] The configuration value for DLA-BRAMAC has the form of (Qvec1+Qvec2, Cvec, Kvec), where Qvec1 and Qvec2 are the numbers of output features computed by DSP and BRAMAC, respectively. 
    \end{tablenotes}
  
  \end{threeparttable}
  
  \label{DLA_BRAMAC_Config}
\end{table*}

\begin{figure*}
    \centering
    \includegraphics[width=1\linewidth]{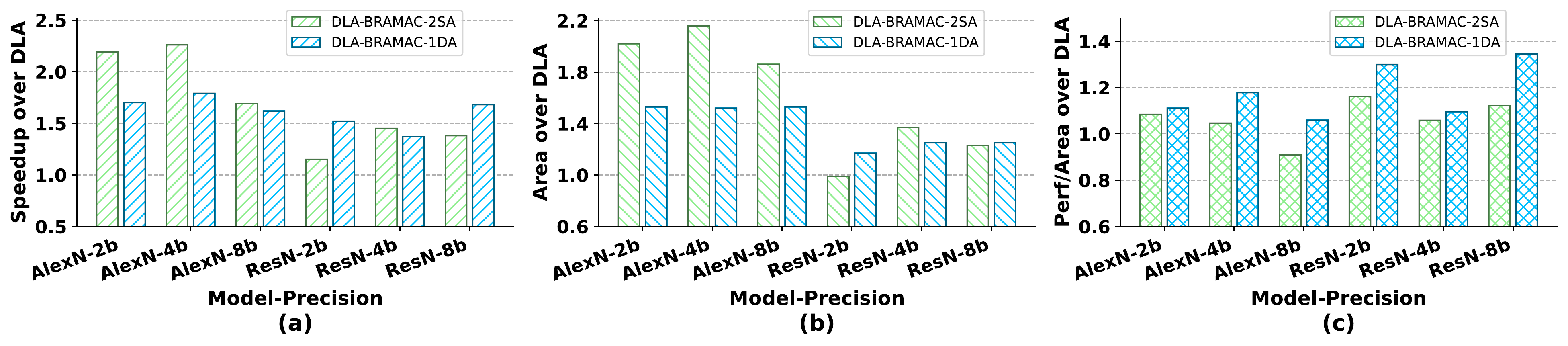}
    \caption{Comparison between DLA and DLA-BRAMAC for accelerating AlexNet and ResNet at different precisions: a) performance, (b) utilized DSP-plus-BRAM area, (c) performance per area. }
    \label{DLA_BRAMAC_Perf}
\end{figure*}

\subsection{Case Study: Employ BRAMAC to Intel's DLA} 
To demonstrate the feasibility of BRAMAC for tiling-based DNN inference with non-persistent weight storage, we employ BRAMAC to Intel's Deep Learning Accelerator (DLA) \cite{DLA_old, DLA_new} and develop a cycle-accurate simulator to model DLA in both the baseline FPGA and the enhanced FPGA with BRAMAC (which we call DLA-BRAMAC). The original DLA is designed to accelerate convolutional neural networks (CNNs) as shown in Fig. \ref{dla_bramac_architecture}(a). It has a processing element (PE) array organized in a 1D systolic structure, a stream buffer to store input and output features, and a filter cache to store weights. It can be parameterized by Cvec, Qvec, and Kvec which represent the computation parallelism per cycle in input depth, output width dimension, and output depth, respectively as illustrated in Fig. \ref{dla_bramac_architecture}(b). 
For DLA-BRAMAC, the stream buffer can send different input features to the PE array and the BRAMAC-based filter cache simultaneously as shown in Fig. \ref{dla_bramac_architecture}(c). In this way, BRAMAC can complement the PE array to calculate different outputs along the Qvec dimension.  

Similar to the approach used in the original DLA \cite{DLA_old}, we conduct design space exploration to find the optimal DLA and DLA-BRAMAC configurations (i.e., Cvec, Qvec, and Kvec) for two popular CNN models: Alexnet and ResNet-34. 
Our analytical model is set to optimize the target function $perf*(perf/area)$ to balance performance and area cost. 
It assumes that all multipliers are implemented using DSPs, and each DSP can pack one 8-bit, two 4-bit, or four 2-bit multiplications using the DSP-packing technique in \cite{dspPacking}.
For area modeling, we use the DLA area model from \cite{DLA_old} to estimate the number of DSPs and BRAMs required for a specific configuration. We ignore the number of ALMs in our area modeling since they are mainly used to implement non-compute-intensive operations and are expected to be similar in DLA and DLA-BRAMAC. 
To evaluate the performance, our cycle-accurate simulator accounts for the latency associated with the MAC2 computation and the dummy array's accumulator readout. 
Note that BRAMAC's eFSM can effectively pipeline adjacent MAC2 operations to hide the latency of the weight copy, except for the first MAC2 of every CNN layer where an additional 2 cycles are required to start the initial weight copy. 
However, this overhead is negligible given that each CNN layer takes thousands of cycles to complete.

Table \ref{DLA_BRAMAC_Config} summarizes the optimal configuration for each (accelerator, model, precision) case. The performance and utilized DSP-plus-BRAM area of DLA-BRAMAC, normalized to those of DLA, are shown in Fig. \ref{DLA_BRAMAC_Perf}. The utilized DSP-plus-BRAM area is calculated based on the area overhead of BRAMAC and the area model from \cite{TILT}. On average, compared to the baseline DLA for AlexNet, employing BRAMAC-2SA/BRAMAC-1DA achieves 2.05$\times$/1.7$\times$ speedup at the cost of 2.01$\times$/1.52$\times$ DSP-plus-BRAM area, giving 1.01$\times$/1.12$\times$ performance gains per utilized area. For ResNet-34, employing BRAMAC-2SA/BRAMAC-1DA achieves a lower speedup of 1.33$\times$/1.52$\times$ on average at the cost of 1.2$\times$/1.22$\times$ DSP-plus-BRAM area, which corresponds to 1.11$\times$/1.25$\times$ performance gains per utilized area. The larger DSP-plus-BRAM area is mainly attributed to more BRAM usage for computation and BRAMAC's area overhead.

In general, BRAMAC-2SA and BRAMAC-1DA achieve higher speedup for AlexNet compared to ResNet-34 as shown in Fig. \ref{DLA_BRAMAC_Perf}(a). 
This is because that BRAMAC is better at supporting a higher Kvec that allows the same input feature to be multiplied by many kernels. 
The early and most compute-intensive residual blocks of ResNet-34 only have an output channel depth of 64, while the first convolution layer of AlexNet has an output channel depth of 96. 
The latter gives more freedom for DLA-BRAMAC to optimize its configuration with high vectorization efficiency. 
However, a higher speedup for AlexNet comes with a larger utilized area as illustrated in Fig. \ref{DLA_BRAMAC_Perf}(b). 
Comparing the two BRAMAC variants, BRAMAC-2SA has a lower performance gain per utilized area for all model-precision combinations as observed from Fig. \ref{DLA_BRAMAC_Perf}(c). 
Although the MAC throughput of BRAMAC-2SA is slightly improved over BRAMAC-1DA, it has 2$\times$ BRAM area overhead compared to BRAMAC-1DA. 
While our results more than justify the area overhead of BRAMAC, we expect higher gains for a DNN accelerator that is: (1) purpose-built around the capabilities of BRAMAC, and (2) used to accelerate DNNs with more matrix multiplications such as transformers \cite{attention}---we will work on both aspects in the future.

\section{Conclusion} \label{Conclusion}
This paper proposes BRAMAC, a compute-in-BRAM architecture for MAC on FPGAs. To the best of our knowledge, BRAMAC is the first compute-in-BRAM architecture that: (1) adopts a hybrid bit-serial \& bit-parallel dataflow to support variable-precision MAC using 2's complement representation, (2) computes in a separate dummy array which improves the main BRAM array's utilization efficiency, (3) employs an embedded finite-state machine to free up the main BRAM ports during in-memory computation. The two proposed variants, BRAMAC-2SA/BRAMAC-1DA, boost the peak MAC throughput of a large Arria 10 FPGA by 2.6$\times$/2.1$\times$, 2.3$\times$/2.0$\times$, and 1.9$\times$/1.7$\times$ for 2-bit, 4-bit, and 8-bit precisions, respectively at the cost of 6.8\%/3.4\% increase in FPGA core area. 
BRAMAC also improves the BRAM utilization efficiency by 1.3$\times$ and 1.1$\times$ compared to two recent compute-in-BRAM architectures, CCB and CoMeFa, respectively while significantly outperforming both architectures on matrix-vector multiplications. 
Combining BRAMAC-2SA/BRAMAC-1DA with Intel's DLA, a tiling-based DNN accelerator, an average speedup of 2.05$\times$/1.7$\times$ and 1.33$\times$/1.52$\times$ can be achieved for AlexNet and ResNet-34, respectively. 
With its ability to support both persistent and tiling-based DNN acceleration, BRAMAC has the potential to be a highly practical and valuable addition to future AI-optimized FPGAs.



\clearpage 
\bibliographystyle{IEEEtran}
\bibliography{references}

\end{document}